\documentclass[aps,amsmath,amssymb,12pt,titlepage,preprint,nofootinbib]{revtex4-2}
\usepackage[dvipdfmx]{graphicx}
\usepackage[breaklinks,colorlinks=true]{hyperref}
\usepackage{booktabs} 
\usepackage{bm}
\usepackage{siunitx}

\newcommand{\be}{\begin{equation}}
\newcommand{\ee}{\end{equation}}

\newcommand{\nl}{\notag\\}

\newcommand{\bx}{\bm{\mathrm{x}}}
\newcommand{\bb}{\bm{\mathrm{b}}}
\newcommand{\bX}{\bm{X}}
\newcommand{\bY}{\bm{Y}}
\newcommand{\calF}{\mathcal{F}}
\newcommand{\calD}{\mathcal{D}}

\newcommand{\calX}{\mathcal{X}}
\newcommand{\calY}{\mathcal{Y}}

\newcommand{\TU}{\mathrm{TU}}

\newcommand{\FEoS}{F_\text{EoS}}

\newcommand{\Nseg}{N_{\text{seg}}}
\newcommand{\NStar}{N_\text{Star}}

\begin{document}

\title{Topological Uncertainty for Anomaly Detection in the Neural-network EoS Inference with Neutron Star Data}

\author{Kenji Fukushima}
\email{fuku@nt.phys.s.u-tokyo.ac.jp}
\affiliation{Department of Physics, The University of Tokyo,
  7-3-1 Hongo, Bunkyo-ku, Tokyo 113-0033, Japan}

\author{Syo Kamata}
\email{skamata11phys@gmail.com}
\affiliation{Department of Physics, The University of Tokyo,
  7-3-1 Hongo, Bunkyo-ku, Tokyo 113-0033, Japan}

\begin{abstract}
   We study the performance of the Topological Uncertainty (TU) constructed with a trained feedforward neural network (FNN) for Anomaly Detection.
   Generally, meaningful information can be stored in the hidden layers of the trained FNN, and the TU implementation is one tractable recipe to extract buried information by means of the Topological Data Analysis.
   We explicate the concept of the TU and the numerical procedures.  Then, for a concrete demonstration of the performance test, we employ the Neutron Star data used for inference of the equation of state (EoS).  For the training dataset consisting of the input (Neutron Star data) and the output (EoS parameters), we can compare the inferred EoSs and the exact answers to classify the data with the label $k$.
   The subdataset with $k=0$ leads to the normal inference for which the inferred EoS approximates the answer well, while the subdataset with $k=1$ ends up with the unsuccessful inference.  Once the TU is prepared based on the $k$-labled subdatasets, we introduce the cross-TU to quantify the uncertainty of characterizing the $k$-labeled data with the label $j$.  The anomaly or unsuccessful inference is correctly detected if the cross-TU for $j=k=1$ is smaller than that for $j=0$ and $k=1$.  In our numerical experiment, for various input data, we calculate the cross-TU and estimate the performance of Anomaly Detection.  We find that performance depends on FNN hyperparameters, and the success rate of Anomaly Detection exceeds $90\%$ in the best case.
   We finally discuss further potential of the TU application to retrieve the information hidden in the trained FNN\@. 
\end{abstract}

\maketitle

\section{Introduction}

Understanding the properties of matter at high baryon density based on the first-principles theory, namely, quantum chromodynamics (QCD), remains a fundamental and unresolved challenge in nuclear physics~\cite{Baym:2017whm,Andronic:2017pug,Fischer:2018sdj,Aarts:2023vsf,Fukushima:2025ujk}. 
Although significant progress has been made in mapping the QCD phase structure at high temperature and low baryon density, where the equation of state (EoS) is accessible in numerical lattice-QCD simulations~\cite{Bollweg:2022fqq,Borsanyi:2023wno}, the cold and dense regime continues to pose severe computational difficulty called the sign problem, which can be evaded in analogous systems~\cite{Son:2000xc,Brandt:2022hwy,Abbott:2023coj,Braguta:2016cpw,Iida:2022hyy}.
Yet, the EoS of cold and dense matter is essential for identifying exotic phases, such as inhomogeneous states~\cite{Buballa:2014tba}, the quasi-long-range order or the moat regime~\cite{Lee:2015bva,Hidaka:2015xza,Pisarski:2021qof}, quarkyonic matter~\cite{McLerran:2007qj}, color superconducting states~\cite{Alford:2007xm}, and so on, with implications for phenomenological modeling of heavy-ion collisions and Neutron Star structures.
In particular, in the latter context of astrophysics, recent x-ray and gravitational-wave observations provide constraints on Neutron Star properties, thereby informing the EoS of dense matter~\cite{Riley:2019yda,*Riley:2021pdl,Miller:2019cac,*Miller:2021qha,LIGOScientific:2017vwq,*LIGOScientific:2018cki}.

Model-independent approaches have been developed with machine learning techniques, i.e., the Bayesian inference~\cite{Steiner:2010fz,Ozel:2015fia, Bogdanov:2016nle} and the deep neural networks (NNs)~\cite{Fujimoto:2017cdo,Fujimoto:2019hxv,Fujimoto:2021zas,Fujimoto:2024cyv,Ferreira:2019bny, Morawski:2020izm, Traversi:2020dho, Krastev:2021reh, Soma:2022qnv, Thete:2022drz, Farrell:2022lfd, Soma:2022vbb, Ferreira:2022nwh, Goncalves:2022smd, Chatterjee:2023ecc, Krastev:2023fnh, Farrell:2023ojk, Zhou:2023cfs, Carvalho:2023ele, Carvalho:2024kgf,Patra:2025xtd}.
Specifically, feedforward neural networks (FNNs) have shown promising performance in classification and inference tasks in general.  Traditionally, plausible effective models are assumed on the physics side, and a ``forward process'' computes the EoSs from these models, rejecting inconsistent ones.  The inverse problem of inferring the EoS is nothing but a ``backward process'' to find the most likely EoS for given experimental input.

In the previous studies developed by the present authors~\cite{Fujimoto:2017cdo,Fujimoto:2019hxv,Fujimoto:2021zas,Fujimoto:2024cyv}, the EoS inference process with FNNs, which we call the FNNEoS, was formulated and has been improved.  The crucial point is that there is one-to-one correspondence between the EoS (i.e., the pressure, $p$, as a function of the energy density, $\epsilon$) and the $M$-$R$ relation (i.e., the distribution of Neutron Stars with mass, $M$, and radius, $R$)~\cite{1992ApJ...398..569L}.  To address the motivation in the present study, let us symbolically explain how the FNNEoS works below.  For the training dataset, $\calD^\text{train} = \{ \bX^{\text{train}(i)}, \bY^{\text{train}(i)} \}_i$, a FNN mapping,
\begin{equation}
  F ~:~ \bX \mapsto \widehat\bY \,,
\end{equation}
is trained, where $\widehat{\bY}$ is a prediction to approximate $\bY^{\text{train}(i)}$ in response to $\bX=\bX^{\text{train}(i)}$.
In the FNNEoS architecture, $\bX$'s characterize the $M$-$R$ relation, and $\bY$'s are the EoS parameters, such as a set of sound velocities at different points of $\epsilon$.
Then, the training optimizes the FNNEoS parameters to make $\widehat{\bY}^{(i)}$ as close to $\bY^{\text{train}(i)}$ as possible for randomly generated EoSs.  Once the training is completed, $F(\cdot)$ can provide us with the most likely EoS corresponding to the real experimental data of Neutron Stars.  This is the general scheme of the supervised learning, not limited to QCD and Neutron Star physics. 

\begin{figure}
  \centering
  \includegraphics[width=0.47\textwidth]{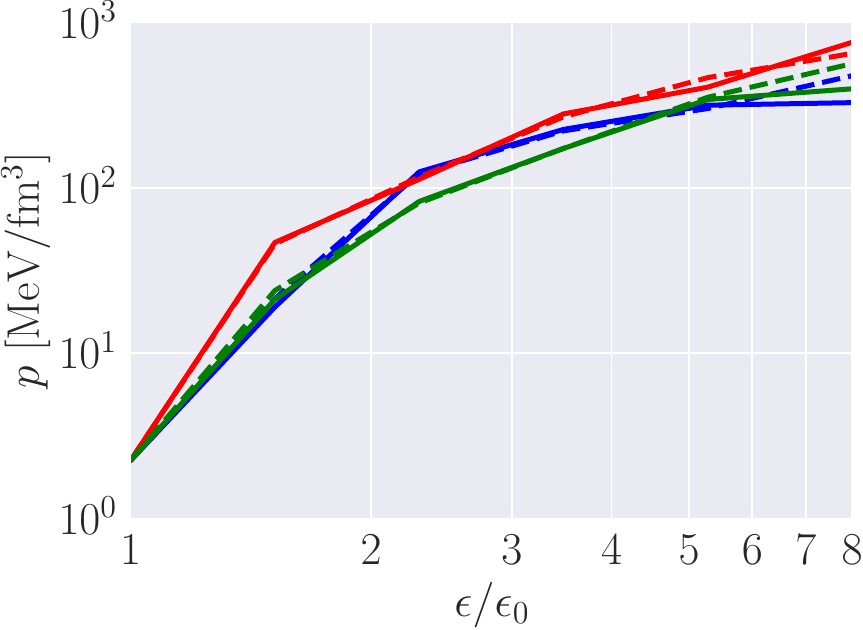}
  \hspace{1em}
  \includegraphics[width=0.47\textwidth]{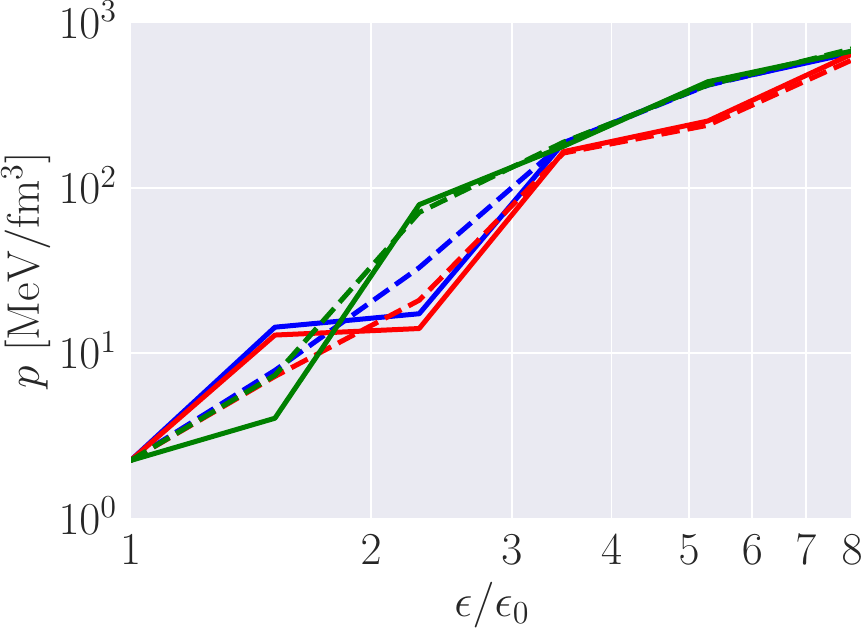}
  \caption{Examples of the successful (left) and the unsuccessful (right) results of the EoS inference.  The dashed lines represent the answers corresponding to $\bY^{\text{train}(i)}$, while the solid lines represent the predictions corresponding to $\widehat{\bY}^{(i)}$.  The energy density along the horizontal axis is given in the unit of $\epsilon_0=150\,\text{MeV/fm$^3$}$ at saturation density of nuclear matter.}
  \label{fig:example}
\end{figure}

In this paper, we discuss the following two questions which are related to each other.
\begin{enumerate}
    \item The trained FNNs contain more information in the hidden layers than that accessed in the output layer only.  So far, in the literature, there are not many discussions for analyzing the bulk properties of the trained FNNs. 
    \item Even if the output from the FNNs is identical, the bulk response including the hidden layers may be different.  The bulk properties could characterize the learning quality to distinguish the successful and unsuccessful applications of the FNNs.
\end{enumerate}
In the present work, we often refer to the second question as Anomaly Detection.  To give intuitive explanations, we shall return to the Neutron Star problem.  Figure~\ref{fig:example} shows the successful (left panel) and the unsuccessful (right panel) examples of the EoS inference using the trained FNNEoS\@.  In this test case, we use some mock dataset (or a validation dataset) $\{\bX^{(i)}, \bY^{(i)}\}_i$ and compare inferred $\widehat{\bY}^{(i)}$ (solid lines) and given $\bY^{(i)}$ (dashed lines).  As seen from apparent difference in two panels of Fig.~\ref{fig:example}, Anomaly Detection is easy if the answers, $\bY^{(i)}$'s, are known.  The same color represents the same $i$, and in the successful case (left), the solid and dashed lines with the same color almost coincide.  In the unsuccessful case (right), on the other hand, the solid and dashed lines show sizable deviations.  Now, the ultimate mission of Anomaly Detection is to judge whether the inference result $\widehat{\bY}$ for the input $\bX$ belongs to the normal class (successful inference) or not without knowing $\bY$.
In general, the unsupervised Anomaly Detection is a challenging problem in science for not only avoiding malfunctions but also discovering new phenomena~\cite{10.5120/13715-1478,9439459,10.1145/3439950}.  In accelerator science, hundreds of millions of events are stored, among which signatures for new science may or may not be contained, and such new science should be detected as anomaly, not belonging to the normal class.

In this work, we will focus on a method of Anomaly Detection based on the Topological Data Analysis (TDA) applied to the trained FNNs, which is nowadays referred to as the Topological Uncertainty (TU)~\cite{lacombe2021}; for TDA reviews, see, e.g., Refs.~\cite{2020NatRP...2..697C,chazal2021introductiontopologicaldataanalysis,Leykam:2022ejk}.
Originally, the TU was proposed to resolve the undesirable behavior of the FNNs for the classification problem~\cite{gebhart2019characterizingshapeactivationspace,lacombe2021}.
Let us consider the following task as a simple example.  The trained FNN is designed to classify something with the output, $0\le y\le 1$, for a given input $x$.  If $x$ is out of the region covered by the training data, one may expect $y\simeq 0.5$, but sometimes the trained FNN may lead to $y$ close to either $0$ or $1$ like the one-hot vector.  In other words, for classification problems, the distance of $y$ from the edging values, $0$ and $1$, may have nothing to do with the likeliness of $y=0$ or $1$.
For science analyses, this is problematic.  The idea of the TU is to diagnose the homological response of the FNN using a TDA pattern.  For this, a filtration is incorporated into the weights of the trained FNN and the maximum spanning tree (MST) of the network graph should be treated.  Importantly, the persistence diagram in the TU construction enjoys stability with respect to small perturbations onto the FNN~\cite{CohenSteiner:2007Stability}.  The TU is complementary to conventional out-of-distribution indicators such as maximum-softmax confidence, out-of-distribution detector for neural networks, and energy-based scores~\cite{Hendrycks:2017,Liang:2018ODIN,Liu:2020Energy}.
Thus, our analysis aligns with neural-persistence style measures and differentiable topology layers~\cite{Rieck:2019NeuralPersistence,Gabrielsson:2020TopologyLayer,Moor:2020TopologicalAE}.

This paper aims to apply the TDA to the neural-network graphs to extract information in the hidden layers.  Although the TU has not been well investigated in a wide range of context, it has a lot of potential for further applications.  We will conduct a numerical experiment to measure the performance of Anomaly Detection.  The quantified performance depends on various hyperparameters of the FNNs as well as our definition of the normal and anomalous inferences.  Introducing a tolerance parameter, $\delta$, that specifies a threshold to bound anomaly, we will see the $\delta$-dependence of the training quality and the Anomaly Detection rate.  The present study is the first successful demonstration of using the TU for Anomaly Detection.

\section{Methodology}

We make an overview of the construction of the TU within the FNN framework in the general context according to Refs.~\cite{gebhart2019characterizingshapeactivationspace,lacombe2021}.

\subsection{Feedforward Neural-Network Graphs}
\label{sec:FNN}

\begin{figure}
    \centering
    \includegraphics[width=0.99\textwidth]{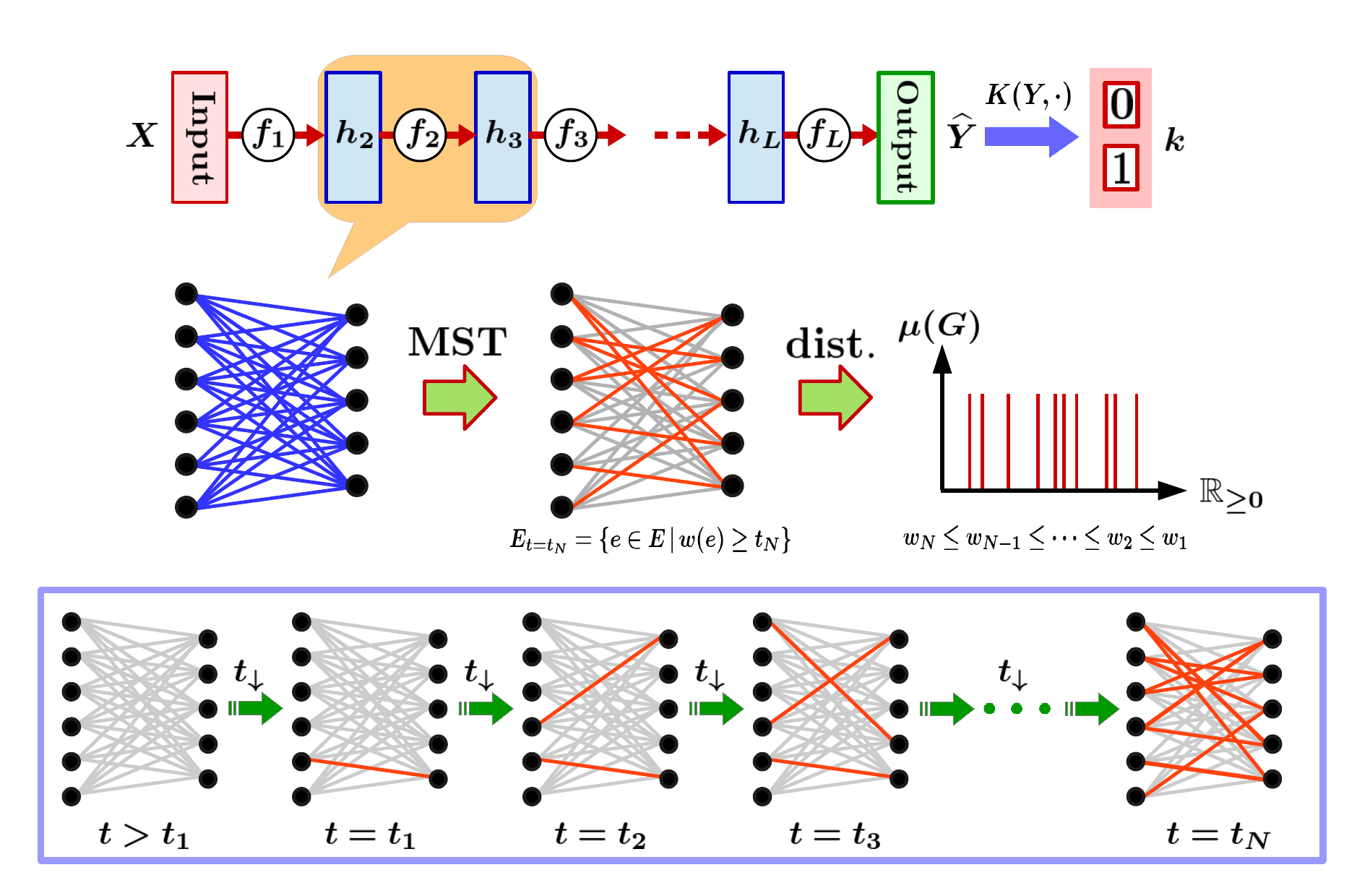}
    \caption{Schematic figure to illustrate the way how to construct the TU from the FNN framework.  See the text for detailed explanations.}
    \label{fig:schematic}
\end{figure}

A generic FNN, denoted by $F(\cdot)$, is defined as a mapping from $\bx_1 \in {\mathbb R}^{d_1}$ to $\bx_{L+1} \in {\mathbb R}^{d_{L+1}}$, where $\bx_1$ and $\bx_{L+1}$ form the input and output layers, respectively, and $\{\bx_2,\dots,\bx_{L-1}\}$ are placed on the $(L-2)$ hidden layers.  Here, $d_\ell \in {\mathbb N}$ represents the dimension of $\bx_\ell$.  We express a mapping from the $\ell$-th to the $(\ell+1)$-th layer by $f_\ell$, i.e.,
\be
  f_\ell \;:\; \bx_\ell\, \mapsto\, \bx_{\ell+1} = {\bm\sigma}_\ell (W_\ell \cdot \bx_\ell + \bb_\ell)\,.
\ee
We note that ${\bm\sigma}_\ell : {\mathbb R}^{d_{\ell+1}} \to {\mathbb R}^{d_{\ell+1}}$ is an activation matrix between two adjacent layers.  For the TU, the weight matrices, $W_\ell\in \mathbb{R}^{d_\ell\times d_{\ell+1}}$, play an essential role as explained below, while the bias vectors, $\bb_\ell \in \mathbb{R}^{d_{\ell+1}}$, will not be used.  In the most cases, the $i$-th unit, $\sigma_\ell^{(i)}$, takes only a single argument of the $i$-th unit of $W_\ell\cdot \bx_\ell + \bb_\ell$.  The FNN is a composite mapping as represented as
\be
  F := f_{L} \circ f_{L-1} \circ \cdots \circ f_2 \circ f_{1}\,.
  \label{eq:F_f}
\ee
To compute the TU below, let us suppose that $F$ has already been trained, that is, $W_\ell$, $\bb_\ell$, and ${\bm\sigma}_\ell$ are all fixed.  In practice, $F$ is designed to solve a classification problem from the input $\bX$ to the output $\widehat{\bY}$.  Moreover, we introduce a label function, $K(\bY,\cdot)$ with the given answer $\bY$, to compute a probability of having the label $k \in \{0,1\}$ in response to $\bX$ through $\widehat{\bY}$; see an upper part in the schematic figure in Fig.~\ref{fig:schematic}.

\subsection{Constructing the Topological Uncertainty}
\label{sec:TU}

The TU is a numerical value computed from $W_\ell$ in the trained FNN in order to characterize how $W_\ell\cdot\bx_\ell$ are activated for each input data and how this activation pattern is related to the zero-dimensional persistent homology.  We shall introduce a filtration for the analysis of the persistent homology.  Generally, the filtration, $\calF$, is expressed as
\be
  \calF_{t \in {\mathbb R}} := \{ x \in X \, | \, f(x) \ge t \} \, \subseteq X\,.
  \label{eq:F_filtr}
\ee
Here, $X$ is a topological space and $f: X\to\mathbb{R}$ is a real-valued continuous function.  We note that $\calF_{+\infty} = \{ \emptyset \}$ and $\calF_{-\infty} = X$.

As seen in Eq.~\eqref{eq:F_f}, $F$ is composed from $f_\ell$ that connects two layers.  Now, we use the language of graph theory for convenience, and we regard each $f_\ell$ as a bipartite graph.  Below, we will focus on one bipartite graph associated with $f_\ell$.  In terms of the graph theory notation, this bipartite graph is written as $G = (V,\,E)$ where $V$ and $E$ are the vertices and the edges, respectively (where the layer index, $\ell$, is omitted for notational brevity).  Then, $V$ corresponds to the neurons in $f_\ell$.  The number of the edges is $|V| = d_\ell + d_{\ell+1}$ and the number of the edges is $|E| = d_\ell \times d_{\ell+1}$.

To introduce the filtration, we define a weight function, $w$, from an edge to a real number as\footnote{For a comparison with neural persistence, see Ref.~\cite{lacombe2021}.}
\be
  w \;:\; e \mapsto | W_\ell(i,j) \bx_\ell(i) |\,,
\ee
where $\bx_\ell^{(i)}$ and $W_\ell^{(i,j)}$ denote the $i$-th unit of $\bx_\ell$ and the ($i$-$j$)-th element of $W_\ell$, respectively, with $e$ an edge connecting the $i$-th neuron on the $\ell$-th layer and the $j$-th neuron on the $(\ell+1)$-th layer.  Using this function, let us introduce a filtration or a filtrated subgraph, $G_t: = (V,\,E_t) \subseteq G$, with
\be
  E_t := \{ e \in E \, | \, w(e) \ge t \}\,.
  \label{eq:def_Gt}
\ee
This filtration in Eq.~\eqref{eq:def_Gt} can be translated to the common TDA procedure.  The neurons are the data points, (the inverse of) $w(e)$ corresponds to the distance between the neurons, and (the inverse of) $t\in [0,+\infty)$ corresponds to the radius of circles around these data points.

For the computation of the TU, we need to find the maximum spanning tree (MST), which corresponds to a persistence diagram of the zero-dimensional homology, $H_0$, in the TDA language.
For a given set of neurons in $f_\ell$, $V=\{\bx_\ell,\, \bx_{\ell+1}\}$, we consider a filtration from $t=+\infty$ and then decrease $t$.  For this initial condition, obviously, $\dim H_0(G_{+\infty}) = |V|$ for $G_{+\infty}=(V,\, E_{+\infty})$.  We define $t_1$ and $e_1$ such that $|E_{t_1}| - |E_{t_1^+}|=1$, $w(e_1)=t_1$, and $\dim H_0(\tilde{G}_{t_1})=|V|-1$ for $\tilde{G}_{t_1}=(V,\, \tilde{E}_{t_1})$ with $\tilde{E}_{t_1}=\{e_1\}$.  We repeat this procedure to find $t_n$ and $e_n$ such that $|E_{t_n}| - |E_{t_n^+}|=1$, $w(e_n)=t_n$, and $\dim H_0(\tilde{G}_{t_n})=|V|-n$ for $\tilde{G}_{t_n}=(V,\,\tilde{E}_{t_n})$ with $E_{t_n}=\{e_i\}_{i=1}^{n}$ until $n=N=|V|-1$.  By definition, it is clear that $\dim H_0(\tilde{G}_{t_N})=1$, and thus $\tilde{G}_{t_N}$ is a graph of dense connections with the least number of edges, and this is how we construct the MST graph.
The procedures are schematically summarized in the bottom of Fig.~\ref{fig:schematic}.

Consequently, for a given MST graph $G$ (we omit the tilde), we obtain a sequence of times, that is, $t_1 \ge t_2 \ge \cdots \ge t_{N-1} \ge t_N$ where $t_n = w_n := w(e_n)$.  Therefore, we also obtain a sequence of filtrated graphs from
$G_{t=+\infty}=(V,\,\{\emptyset\})$ with $\dim H_0(G_{+\infty})=|V|$
to
$G_{t<t_N}=G$ with $\dim H_0(G_{t<t_N}) = 1$.
This structure, i.e., the persistence diagram of $H_0$, can be expressed by a probability distribution,
\be
   \mu(G) := \frac{1}{N} \sum_{i=1}^{N} \delta_{w_i}\,,
\ee
where $\delta_{w_i}$ is the Dirac measure defined for a given $x\in X$ as
\be
  \delta_x(Y) =
\begin{cases}
  1 & \text{if $x \in Y$} \\
  0 & \text{otherwise}
\end{cases}
\ee
in a topological space $X$ and a subspace $Y \subseteq X$.

It is important to note that $\mu(G)$ is determined as a function of the input $X$ once the FNN is trained.  When we have two distributions, $\mu = \frac{1}{N} \sum_{i=1}^{N} \delta_{w_i}$ and $\nu = \frac{1}{N} \sum_{i=1}^{N} \delta_{w^{\prime}_i}$, we need to quantify the difference between them by introducing a distance.  For the purpose of the TU, one can employ the $p=1$ Wasserstein distance.  Let us define $\mathrm{Dist}(\mu,\nu)$ as
\be
  \mathop{\mathrm{Dist}}(\mu,\nu) := \frac{1}{N} \sum_{i=1}^N \bigl| w_i - w^{\prime}_i \bigr|\,.
  \label{eq:def_wass1}
\ee
Futhermore, we need a reference probability to classify the topological structure of new data fed into the FNN\@.  For this purpose, one employs the Fr\'{e}chet mean defined from the training sub-dataset, $\calD_k^{\text{train}}$ labeled by $k$, where $k$ is an output in the classification problem as shown in Fig.~\ref{fig:schematic} and its implementation will be explained below.  The averaged probability tagged with $k$ is $\bar{\mu}_k := \frac{1}{N} \sum_{i=1}^{N} \delta_{\bar{w}_{k,i}}$, where
\be
  \bar{w}_{k,i}
  := \frac{1}{|\calD_k^\text{train}|} \sum_{m=1}^{|\calD_k^\text{train}|} w^{(m)}_{k,i}\,.
  \label{eq:def_mubar}
\ee
Here, $w^{(m)}_{k,i}$ denotes $w_i$ as constructed above for the $m$-th data in $\calD_k^{\text{train}}$.  Intuitively, the Fr\'{e}chet mean estimates the average of all probability distributions for data in the training sub-dataset.

Finally, we can define the TU (Topological Uncertainty).  So far, we have focused on $f_\ell$, but now we take a sum over all the layers, $\ell\in\{1,\dots,L\}$.  For one input data, we should find the MST graph, $G_\ell$, and the associated probability distribution, $\nu_\ell(G_\ell)$.  Then,
\begin{equation}
   \mathrm{TU}_k(X) := \frac{1}{L} \sum_{\ell = 1}^L
   \mathop{\mathrm{Dist}} \bigl( \nu^{(\ell)}(X),\, \bar{\mu}^{(\ell)}_k \bigr)\,,
   \label{eq:TUk}
\end{equation}
where $\bar{\mu}^{(\ell)}_k$ is the Fr\'{e}chet mean generated by ${\cal D}^{\rm train}_k$ on $G_\ell$.

\section{Numerical experiment}
We perform the numerical experiment to test the performance of the TU analysis using concrete data in astro-nuclear physics.  We will first explain the physics background in Sec.~\ref{sec:prelim}, then move on to our procedure to compute the TU in Sec.~\ref{sec:procedures}.  Finally, we will discuss the performance test in Sec.~\ref{sec:num_results}.

\subsection{Preparation --- Equation of state and the Neutron Star data}
\label{sec:prelim}

We shall give a brief overview of the physics background.  We will deal with the data to characterize the Neutron Star properties according to the literature~\cite{Fujimoto:2024cyv}.
The task for the FNN to solve is the inference of the EoS parameters, $\bY$, corresponding to the observation, $\bX$.
In the present case of the application, more specifically, $\bX$ refers to the measurement of the Neutron Star, namely, a set of the masses and the radii of $\NStar$ (typically $\sim 15$\,--\,$20$) Neutron Stars, and $\bY$ is a set of EoS parameters, i.e., the density-dependent sound velocity:
\be
  \bX = \{ M_{(m)},\, R_{(m)}\}_{m=1}^{\NStar},
  \qquad
  \bY = \{ c_{s,n}^2 \}_{n=1}^{\Nseg} \in [0.01,0.99]^{\Nseg}\,,
\ee
where $\Nseg$ (typically $\sim 5$) the number of segments to discretize the density relevant to the Neutron Star structures.  For numerical experiments we adopt standard parametrization, i.e., piecewise polytropes~\cite{Read:2008iy}.  We take the lowest energy density as $\epsilon_0 = 150\,\text{MeV}/\text{fm}^3$ that is the energy density at nuclear saturation density and choose the highest energy density as $\epsilon_{\Nseg}=8\epsilon_0$ from the physical boundary condition.  Then, we introduce the discretized energy density points as
\be
  \log\epsilon_n := \log\epsilon_0 + \frac{n}{\Nseg} (\log \epsilon_{\Nseg} - \log \epsilon_0)\,, \quad n \in \{ 1, \cdots, \Nseg \}\,.
\ee
From the inferred sound velocity squared $c_{s,n}^2$ at the energy density $\epsilon_n$, the pressure on the grid is given by
\be
  p_n = p_{n-1} + c_{s,n}^2 (\epsilon_n - \epsilon_{n-1})\,.
  \label{eq:eos.piecewise-pressure}
\ee
There is a mapping relation called the Tolman-Oppenheimer-Volkoff (TOV) equation~\cite{Tolman:1939jz, Oppenheimer:1939ne, 1992ApJ...398..569L}, which connects $\bX$ and $\bY$.  If $\NStar$ and $\Nseg$ are infinitely large, the TOV equation establishes a one-to-one correspondence between $\bX$ and $\bY$.  In this problem, three coupled differential equations are solved:
\be
  \frac{dp}{dr} = - (\epsilon + p) \frac{m_r + 4 \pi r^3 p}{r(r - 2 m_r)}\,,
  \qquad
  \frac{dm_r}{dr} = 4 \pi r^2 \epsilon\,,
  \qquad
  p=p_\text{EoS}(\epsilon)\,.
\ee
Here, $r$ is the distance from the Neutron Star center and the last relation is nothing but the EoS\@.  With a free parameter, $p_c>0$, representing the pressure at the center which differs for different Neutron Stars, the differential equations are integrated from the initial condition, $p(r=0) = p_c$ and $m_r(r=0) = 0$.
The Neutron Star radius, $R$, is determined by the surface condition, $p(r=R)=0$, and then the total mass of the Neutron Star is given by $M=m_r(r=R)$.
In this way, we can draw a line, $(R(p_c),\, M(p_c))$, with a mediator variable $p_c$, or obtain a function $M=M(R)$.  It has been proven that $p=p_\text{EoS}(\epsilon)$ is fully reconstructed from $M=M(R)$.

For the purpose of training the neural network, $F$, which we specifically call the FNNEoS, the training dataset is prepared and the trained network is denoted as $\FEoS$.  The training input-dataset, $\calX$, and the training output-dataset, $\calY$, are defined as
\be
  \calX:= \{\bX^{(i)}\}_{i=1}^{N_\text{data}}\,,
  \qquad
  \calY:= \{\bY^{(i)}\}_{i=1}^{N_\text{data}}\,,
\ee
and the training dataset is represented as
\be
  \calD^\text{train} := \{ \bX^{(i)},\, \bY^{(i)} \}_{i = 1}^{N_\text{data}} \,,
\ee
where $N_\text{data}$ is the size of the training dataset.

The FNNEoS is designed to find the inverse function of the TOV equation.
The FNNEoS is the standard fully connected dense network.
We use the following notation for the structure of the FNNEoS:
\begin{align}
& F_{(h_2,h_3,\cdots,h_{L})} \ : \ \boxed{\text{input}} \rightarrow  \boxed{\text{hidden}_2} \rightarrow \cdots \rightarrow \boxed{\text{hidden}_L} \rightarrow  \boxed{\text{output}}
\nl 
&  d_{\ell \in \{ 2, \cdots, L \}} =  d_{1} \times  h_{\ell}, \quad h_{\ell}\in {\mathbb N}, \quad L \in {\mathbb N} + 1. \label{eq:NN_str}
\end{align}
It should be noted that, unlike a problem explained in Sec.~\ref{sec:FNN}, the FNNEoS does not deal with a classification problem by itself.  Thus, we will later introduce a label function to attach a label to the data.

\subsection{Procedures}
\label{sec:procedures}
To demonstrate the TU analysis, we first introduce the label function to judge whether two EoSs are sufficiently close or not.  Let $\bX\in\calX$ and $\bY\in\calY$ so that $\{\bX,\,\bY\}\in \calD^\text{train}$.  Then, $\widehat{\bY}=\FEoS(\bX)$ is the prediction from the trained $\FEoS$;  if the inference works successfully, $\widehat{\bY}$ is expected to be close to $\bY$.  We would like to measure the distance between $\bY$ and $\widehat{\bY}$ using the $L_\infty$-norm.  For this purpose, we defined the following quantity with the logarithmic difference of the pressure, $p(c_s^2)$, in Eq.~\eqref{eq:eos.piecewise-pressure} as
\be
  \Delta \log p(\bY,\widehat{\bY}) := \sup_{n \in \{ 1, \cdots ,N_\text{trun} \}} \bigl| \log p(y_{(n)}) - \log p(\widehat{y}_{(n)}) \bigr|
  \label{eq:Delta_Log_P}
\ee
with $y_{(n)} \in \bY$ is the sound velocity at the $n$-th segment of the energy density in the training data and $\widehat{y}_{(n)} \in \widehat{\bY}$ is the inferred sound velocity in the prediction data.  The segment index, $n$, runs from 1 (lowest energy density) to the truncation parameter, $N_\text{trun} \in \{ 1, \cdots, \Nseg\}$, to see the cutoff dependence of the highest energy density.  The definition of the label function with a hyperparameter, $\delta\in\mathbb{R}_{>0}$, is
\be
  K_\delta(\bY,\widehat{\bY}) := \begin{cases}
  0 & \text{if $\Delta \log p(\bY,\widehat{\bY}) < \delta$} \\
  1 & \text{otherwise}\,.
\end{cases}\,,
\label{eq:K_func}
\ee
where $\delta$ controls the threshold to regard two EoSs as sufficiently close to each other, which we call the tolerance parameter here.  Thus, we obtain $k=K_\delta(\bY,\widehat{\bY})$, and then the label of $k=0$ means that two EoSs, $\bY$ and $\widehat{\bY}$, can be regarded as consistent, and the label of $k=1$ signifies the failure of the sensible prediction from the FNNEoS\@.

In this way, one can classify all the data in $\calD^\text{train}$ with the label $k$.  This decomposition is symbolically represented as
\be
  \calD^\text{train} = \calD^\text{train}_{k=0} \cup \calD^\text{train}_{k=1} \,,
  \qquad
  \calD^\text{train}_{k=0} \cap \calD^\text{train}_{k=1} = \emptyset\,.
  \label{eq:D_decompose}
\ee
Then, we define the following ratio to measure the learning quality of the FNNEoS, that is, the ratio between the sizes of $\calD_{k=0}^\text{train}$ and $\calD^\text{train}$.
\begin{equation}
    R_\text{LQ}(\delta) := \frac{{|{\cal D}_{k=0}^{\rm train}|}}{|{\cal D}^{\rm train}|} \,.
    \label{eq:R_LQ}
\end{equation}
We note that this ratio depends on the threshold $\delta$.  Although $R_\text{LQ}(\delta)$ is not necessary to construct the TU, we will estimate it to monitor the learning quality for different values of $\delta$.

Here, let us enumerate the procedures for calculating $\TU(\cdot)$ based on $\calD^\text{train}$.  For demonstration, moreover, we prepare a new dataset, $\calD^\text{new}$, independent of $\calD^\text{train}$.  We can compute $\TU_k(\bX^\text{new})$ for $\{\bX^\text{new},\, \bY^\text{new}\} \in \calD^\text{new}$.  This is the TU analyzed result for the Anomaly Detection in $\calD^\text{new}$.  We emphasize, however, that this is not our goal yet;  we should quantify the effectiveness of the TU analysis.  For this purpose, we decompose $\calD^\text{new}$ into $\calD^\text{new}_{k\in\{0,1\}}$, though there is no way to make this decomposition in the practical problem due to the lack of knowledge about $\calY^\text{new}$.  To judge how correctly $\TU_k(\cdot)$ can perform the Anomaly Detection, we consider $\TU_j(X_k\in\calX^\text{new}_k)$.  Recall that the uncertainty is smaller for the smaller TU value.  If $\displaystyle \arg\min_j \TU_j(\bX_k)$ is 0 or 1, respectively, the data is classified to be normal or anomalous, while the answer label is given by $k$.  We introduce the averaged TU to represent the overall performance as
\begin{equation}
  \TU_j(\calX_k^\text{new}) := \frac{1}{|\calX^\text{new}_k|} \sum_{ \bX_k \in \calX_k^\text{new}} \TU_j(\bX_k)\,.
\end{equation}
Since this measures the uncertainty for multiple labels, we call it the \textit{cross-TU} in this work.
Then, we extract the subset of $\calX_k^\text{new}$ in which the TU analysis succeeds in correctly labeling the data with $j$.  That is, the correctly identified subset is
\begin{equation}
  \mathring{\calX}^\text{new}_k := \{ \bX_k \in \calX^\text{new}_k \, | \, \arg \min_{j \in \{0,1\}} \TU_j(\bX_k)  = k \}\,.
  \label{eq:X_correct}
\end{equation}
  
Then, we shall take the following four steps.

\noindent
\paragraph*{\textbf{Step 1}:}
Sample $\{\bX,\, \bY\}\in \calD^\text{train}$ and calculate $\widehat{\bY}=\FEoS(\bX)$.

\noindent
\paragraph*{\textbf{Step 2}:}
For sampled $\bY$ and associated $\widehat{\bY}$, calculate $k=K_\delta(\bY,\widehat{\bY})$ for a fixed $N_\text{trun}$, and attach the label $k$ to the data.
Repeat \textit{\textbf{Step 1}} and \textit{\textbf{Step 2}} until the decomposition of Eq.~\eqref{eq:D_decompose} is completed.  Then, $\calX^\text{train}_k$ and $\calY^\text{train}_k$ are identified from $\calD^\text{train}_k$, respectively.

\noindent
\paragraph*{\textbf{Step 3}:}
Compute the Fr\'{e}chet mean, $\bar{\mu}_k$, using $\calX^\text{train}_k$ and the weights, $W_\ell$, in the trained FNNEoS\@.

\noindent
\paragraph*{\textbf{Step 4}:}
Prepare the test dataset, $\calD^\text{new}$.  
Compute their TU using the Fr\'{e}chet mean obtained in \textit{\textbf{Step 3}}.
The performance or the score of the TU analysis is estimated by the ratio of correctly identified
subset for each $k$; that is,
\begin{equation}
  R_k(\delta) := \frac{|\mathring{\calX}^\text{new}_k|}{|\calX^\text{new}_k|}\,,
  \qquad
  R_\text{tot}(\delta) := \frac{ \sum_{k \in \{0,1\}} |\mathring{\calX}^\text{new}_k| }{ \sum_{k \in \{0,1\}} |\calX^\text{new}_k| }\,,
  \label{eq:TU_score}
\end{equation}
where $\mathring{\cal X}^{\rm new}_{k}$ is defined in Eq.~\eqref{eq:X_correct}.

\begin{table}
  \centering
  \begin{tabular}{cccccccc}
  \toprule
    $d_{1}$ & $d_{L+1}$ & $N_\text{sample}$ & $N_{MR}$ & $\quad \sigma_{\ell<L}\quad$ & $\sigma_{L}$ & Loss func. & Optimizer  \\
  \midrule
    \, 40 \, & 5 & 20 & \num{20000} & ReLU & sigmoid & ${\tt msle}$ & {\tt Adam} \\
  \bottomrule
  \end{tabular}
  \caption{A summary of the adopted Neural Network architecture in the present study.}
  \label{tab:NNEoS}
\end{table}

\begin{table}
  \centering
  \begin{tabular}{c|ccccccccc}
  \toprule
    $F_{(h_2,\cdots,h_\ell)}$  & $F_{(2)}$ & $F_{(2,2)}$ & $F_{(2,2,2)}$ & $F_{(2,2,2,2)}$ & $F_{(4)}$ & $F_{(4,4)}$ & $F_{(4,4,2)}$ & $F_{(4,4,2,2)}$ & $F_{(4,4,4,4)}$ \\
  \midrule  
    Train.\ loss.\ $(\times 10^{-2})$ & $1.89$ & $1.55$ & $1.42$ & $1.36$ & $1.73$ & $1.44$ & $1.35$ & $1.31$ & $1.25$ \\ 
    Val.\ loss.\ $(\times 10^{-2})$ &  $1.89$ & $1.55$ & $1.43$ & $1.36$ & $1.75$ & $1.46$ & $1.38$ & $1.35$ & $1.35$ \\
  \bottomrule     
  \end{tabular}
  \caption{Setup of the training process of the FNNEoS, and the values of the training and the validation loss.
    The notation of $F_{(h_2,\cdots,h_\ell)}$ is introduced in Eq.~\eqref{eq:NN_str}.
    The number of epoch is 400, and the batch size is 250.
    $N_\text{data} = N_\text{sample} \times N_{MR}$.
} 
  \label{tab:loss}
\end{table}

For the training of the FNNEoS, we chose $\NStar = \num{20}$ and $\Nseg = \num{5}$.
We generated $N_{MR}=\num{20000}$ lines from the TOV equation and took $N_\text{sample}=20$ sample points from each $M$-$R$ line, as summarized in Tab.~\ref{tab:NNEoS}.
From the total data whose size is $N_\text{data} = N_\text{sample} \times N_{MR} = \num{400000}$ in the training process,
we used $\num{0.1} \times N_\text{data}$ for the validation process.
We employed the Keras package, and the values of the training and validation losses are listed in Tab.~\ref{tab:loss}.
For the TU analysis, we took $N_\text{tr} = 3$ for the truncation and $\delta = 0.1, 0.2, \cdots, 0.8$ for the threshold.
We prepared new datasets, $\calD^\text{new}_k$, with $|\calD^\text{new}_k|=500$ for each $k\in \{0,1\}$.

\subsection{Results}
\label{sec:num_results}

\begin{figure}
  \centering
  \includegraphics[width=0.5\textwidth]{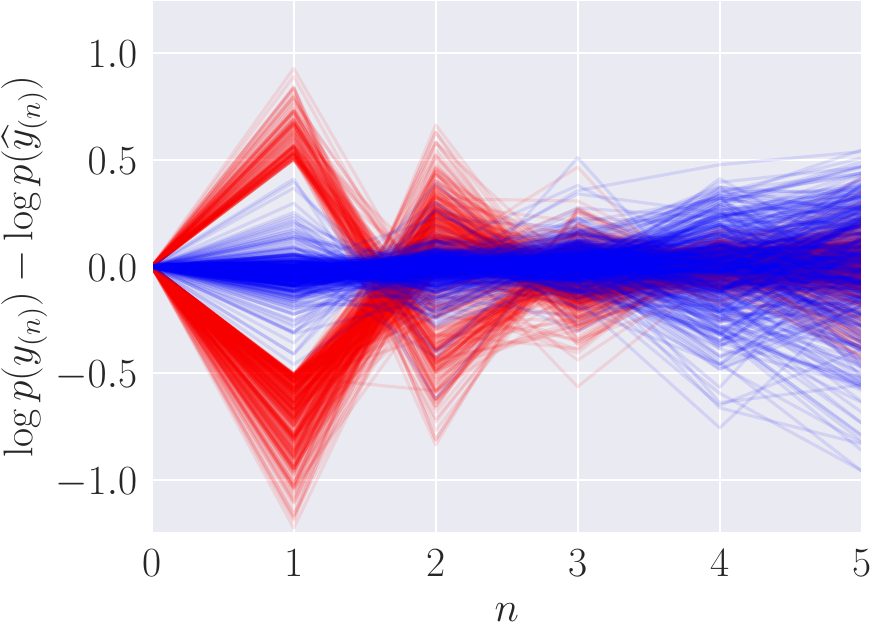}
  \caption{Logarithmic separation between the predicted EoSs and the exact answers quantified by $\log p(y_{(n)}) - \log p(\widehat{y}_{(n)})$.  The inferred EoSs result from $F_{(4,4,2,2)}$ and the tolerance for the label, $k\in \{0,1\}$, is chosen as $\delta = 0.5$.  
  The blue and red lines represents the $k=0$ and $k=1$ samples, respectively, and 500 samples are plotted for each $k$.}
  \label{fig:logp_diff}
\end{figure}

First, we look at the results in the training part.  We compute the predicted EoSs from $\calD^\text{train}$ and then attach the answer labels, $k\in \{0,1\}$, according to Eqs.~\eqref{eq:Delta_Log_P} and \eqref{eq:K_func}.  The example of labeled data for $F_{(4,4,2,2)}$ is shown in Fig.~\ref{fig:logp_diff} where the logarithmic separation between the predicted EoS and the exact answer, $\log p(y_{(n)}) - \log p(\widehat{y}_{(n)})$, is plotted at each energy-density segment.
The $k=0$ samples (blue lines) are classified as correctly inferred EoSs with the tolerance parameter, $\delta=0.5$, while the $k=1$ samples (red lines) are judged as wrongly inferred.  We note that the deviation is estimated up to $n=N_\text{trun}$ in Eq.~\eqref{eq:Delta_Log_P} and we choose $N_\text{trun}=3$ here.  Therefore, the derivations at $n=4$ and $5$ in Fig.~\ref{fig:logp_diff} are completely irrelevant, which explains the spreading behavior of the $k=0$ samples at higher energy-density segments.  Also, naturally, the $k=1$ samples dominantly have the deviations from the exact answers at $n=1$ and $2$.  Interestingly, we see a general tendency that, if the pressure is overestimated (or underestimated) at $n=1$, it is underestimated (or overestimated) at $n=2$.  The same tendency is also manifest in the right panel of Fig.~\ref{fig:example}.

\begin{figure}
  \centering
  \includegraphics[width=0.5\textwidth]{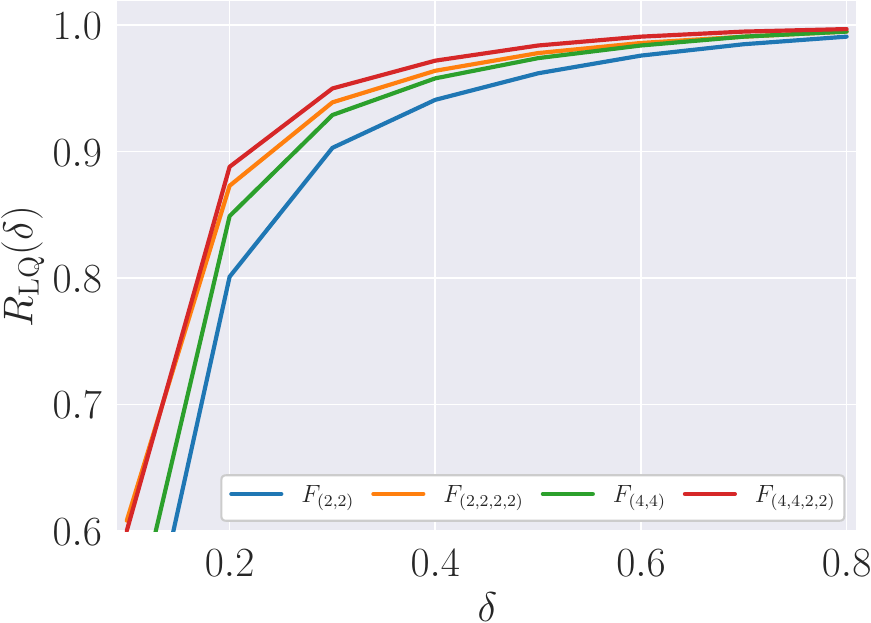}
  \caption{$R_\text{LQ}(\delta)$ (ratio of the $k=0$ data size to the whole data size) for various $F_{(\cdot)}$.}
  \label{fig:R_LQ}
\end{figure}

In Fig.~\ref{fig:R_LQ}, we present the learning quality, $R_\text{LQ}(\delta)$ defined in Eq.~\eqref{eq:R_LQ}, as a function of the tolerance parameter $\delta$.  If the tolerance parameter is larger, more outputs are judged as correct, i.e., $k=0$, and thus $R_\text{LR}$ should increase monotonically with $\delta$.
From Fig.~\ref{fig:R_LQ}, we see that the learning quality is pretty high;
$R_\text{LQ}(\delta>0.2) \gtrsim 0.8$ for all tested $F_{(\cdot)}$.  This means that more than $80\%$ of the outputs belong to correctly inferred EoSs, but $10$--$20\%$ of the outputs are incorrect.  Our task is to automate the Anomaly Detection to identify the incorrect outputs without knowing the exact answers.

\begin{figure}
  \centering
  \includegraphics[width=0.48\textwidth]{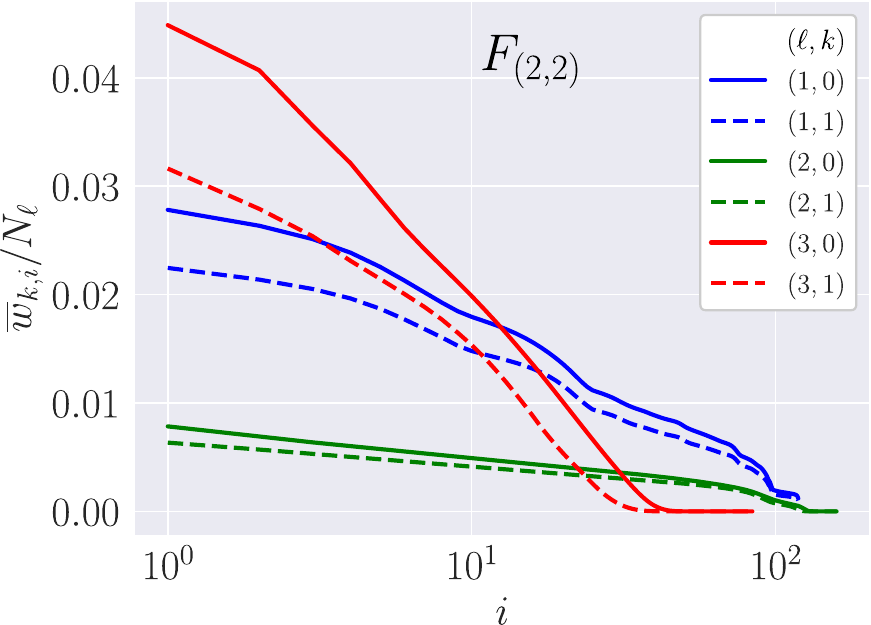}
  \hspace{0.5em}
  \includegraphics[width=0.48\textwidth]{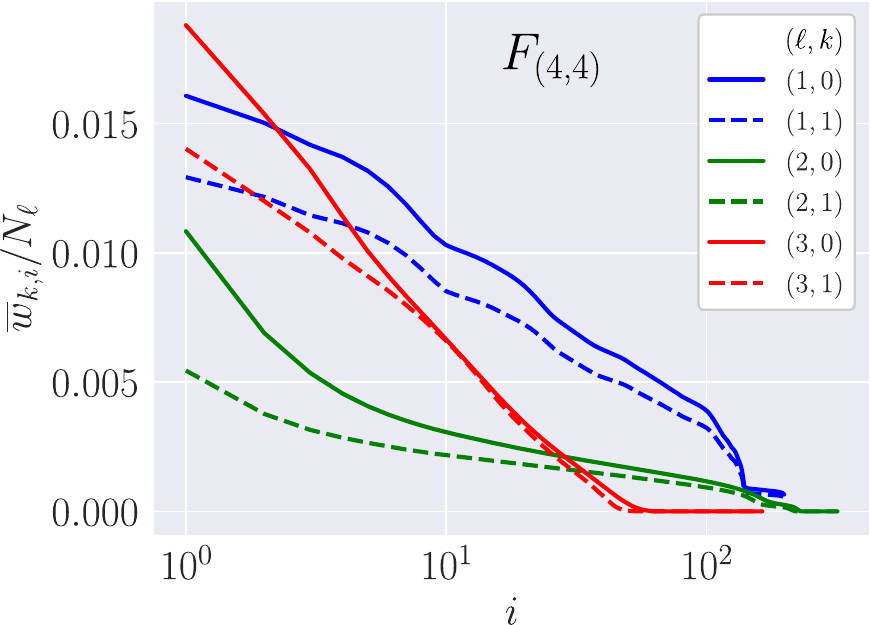}
  \vspace{0.5em}\\
  \includegraphics[width=0.48\textwidth]{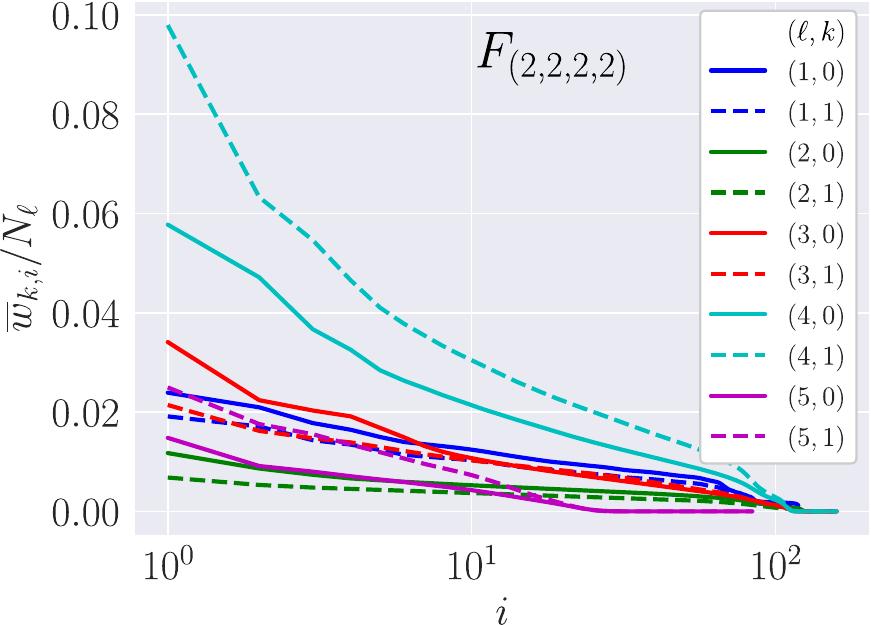}
  \hspace{0.5em}
  \includegraphics[width=0.48\textwidth]{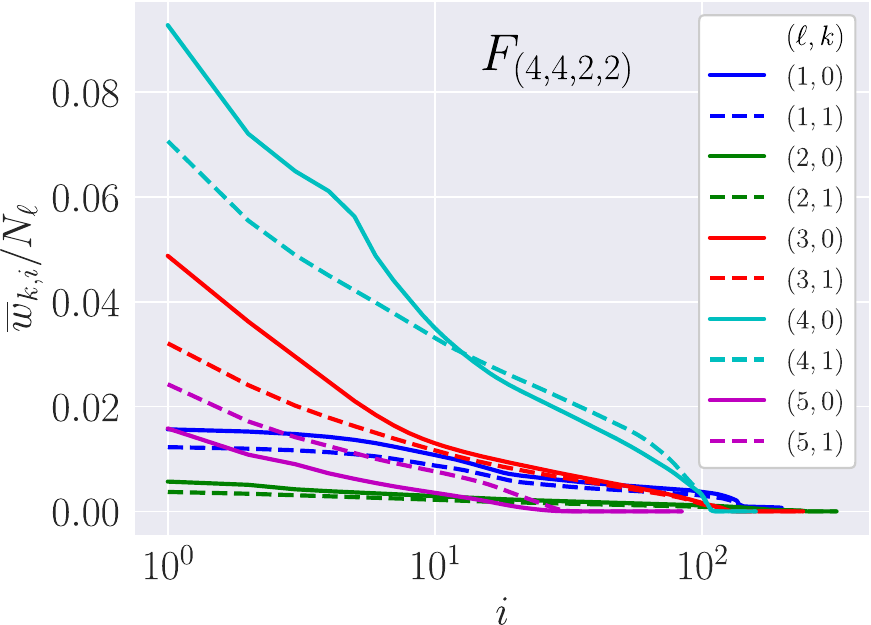}
  \caption{Normalized distribution, $\bar{w}_{k,i}/N_\ell$, in the Fr\'{e}chet mean, $\bar{\mu}_{k,\ell}$, for $k=0$ (solid lines) and $k=1$ (dashed lines).  The normalization is $N_\ell=|V_\ell|-1$ for $f_{\ell\in \{1,\dots,L\}}$.  The distribution is evaluated from $F_{(2,2)}$ (left-top), $F_{(4,4)}$ (right-top), $F_{(2,2,2,2)}$ (left-bottom), and $F_{(4,4,2,2)}$(right-bottom) with the label $k$ for the tolerance $\delta = 0.5$.  In the legend, various choices of $(\ell,k)$ are shown.}
  \label{fig:frechet_means_del05}
\end{figure}

Figure~\ref{fig:frechet_means_del05} shows the distribution, $\bar{w}_{k,i}$, normalized by $N_\ell$ in the Fr\'{e}chet mean, $\bar{\mu}_{k,\ell}$ in Eq.~\eqref{eq:def_mubar} for $k=0$ (solid lines) and $k=1$ (dashed lines).  The normalization is $N_\ell=|V_\ell|-1$ for $f_{\ell\in\{1,\dots,L\}}$.
The panels display the results for $F_{(2,2)}$ (left-top), $F_{(4,4)}$ (right-top), $F_{(2,2,2,2)}$ (left-bottom), and $F_{(4,4,2,2)}$ (right-bottom) with the label $k$ for the tolerance $\delta = 0.5$.

It should be noted that the TU measures the distance between two distributions for $k=0$ and $1$, so a better performance can be expected for the Anomaly Detection if $\bar{w}_{k=0,i}$ and $\bar{w}_{k=1,i}$ appear more separated.  In Fig.~\ref{fig:frechet_means_del05}, therefore, the Anomaly Detection should work better if the solid line and the dashed line with the same color become more widely separated.  It is interesting that the separation can be large for $i\lesssim 10^1$, while it is suppressed for larger $i$ for any case.  We note that, in our convention of the ordering, $\bar{w}_i \ge \bar{w}_{i'}$ for $i< i'$, and the outputs from the FNNEoS may well depend dominantly on the network edges associated with such small $i$.

\begin{figure}
  \centering
  \includegraphics[width=0.48\textwidth]{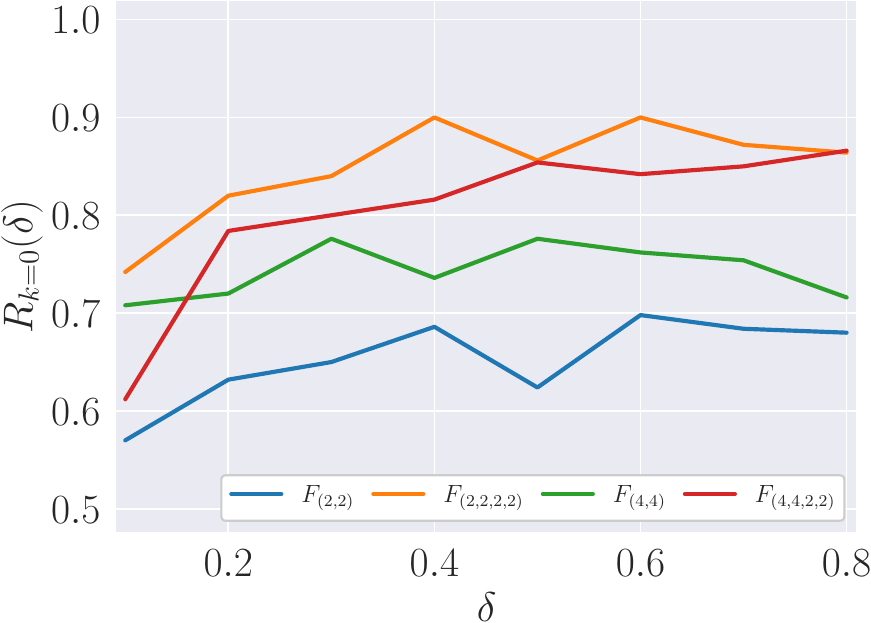}
  \hspace{0.5em}
  \includegraphics[width=0.48\textwidth]{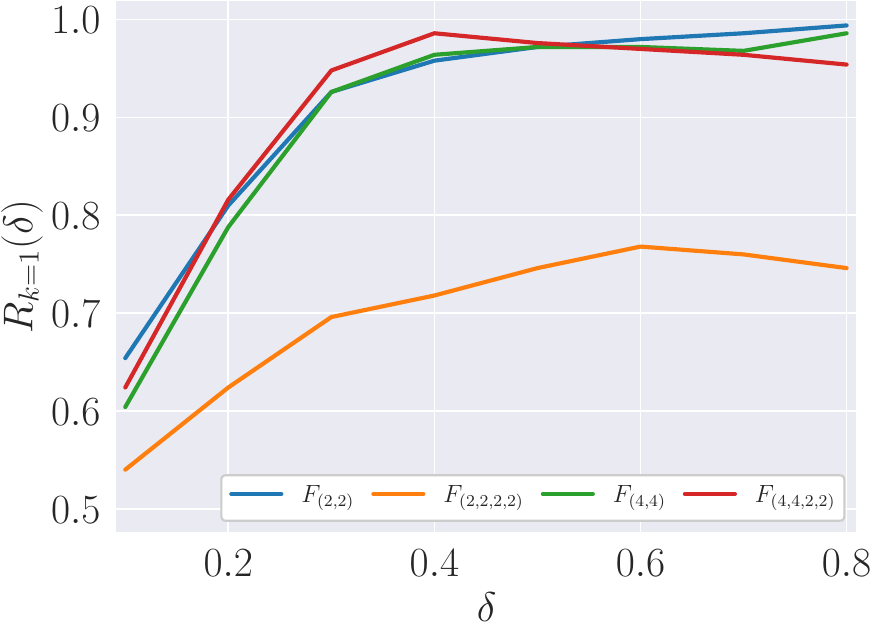}
  \vspace{0.5em}\\
  \includegraphics[width=0.48\textwidth]{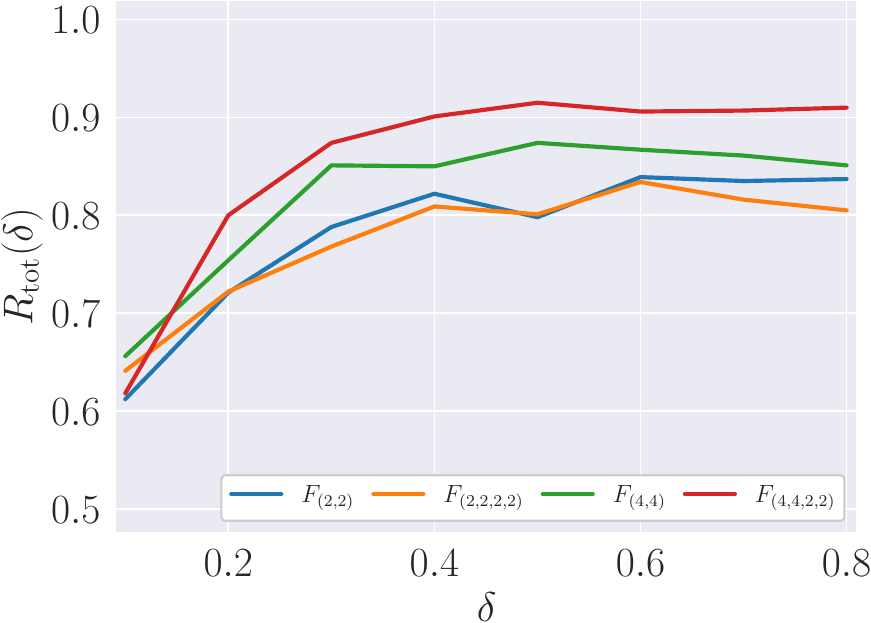}
  \caption{$R_{k=0}(\delta)$ (left-top), $R_{k=1}(\delta)$ (right-top), and $R_\text{tot}(\delta)$ (bottom) for various $F_{(\cdot)}$.}
  \label{fig:score}
\end{figure}

Next, let us discuss the Anomaly Detection performance.  Figure~\ref{fig:score} shows the $\delta$ dependence of $R_{k=0}$ (left-top), $R_{k=1}$ (right-top), and $R_\text{tot}$ (bottom).
If $R_{k=0}$ (and $R_{k=1})$ becomes larger, the correctly (and incorrectly) inferred EoSs are better classified as correct (and incorrect, respectively).  The overall performance is quantified by $R_\text{tot}$.

Generally speaking, $R_\text{tot}$ tends to be improved for a larger-sized FNNEoS, i.e., the neural network with larger depth and width in the hidden layers.  In contrast, once we look into $R_{k=0}$ and $R_{k=1}$ separately, we realize that the performance has more nontrivial structures.
For the case with $k=0$ (correct inference), $R_{k=0}$ starts around $0.6$-$0.8$ at $\delta = 0.1$, and there is no drastic improvement after the saturation around $\delta = 0.2$-$0.3$.  This behavior is qualitatively different from the case with $k=1$.
We see that $R_{k=1}$ rapidly increases in a range of $0.1< \delta \lesssim 0.3$ approaching nearly the unity except for $F_{(2,2,2,2)}$.
To see the dependence on the FNNEoS size, obviously, the performance of $F_{(2,2)}$ is poor in view of $R_{k=0}$ for any $\delta$, while $R_{k=1}$ exhibits performance comparable to others.  This means that, if the neural-network size is small, the TU tends to misjudge the incorrect inference as correct (which lowers $R_{k=0}$), but it is still easy to identify the incorrect inference as incorrect (as indicated by $R_{k=1}\simeq 1$).
One puzzling feature is the behavior of $F_{(2,2,2,2)}$.  As naturally expected, the largest of tested netwroks, i.e., $F_{(4,4,2,2)}$ gives the best performance, and one may think that the second largest one, $F_{(2,2,2,2)}$ should be the second best, but it is not the case.
Strangely, $F_{(2,2,2,2)}$ is superior to $F_{(4,4,2,2)}$ for $R_{k=0}$, but it is terribly poor for $R_{k=1}$ even worse than $F_{(2,2)}$.

\begin{figure}
  \centering
  \includegraphics[width=0.48\textwidth]{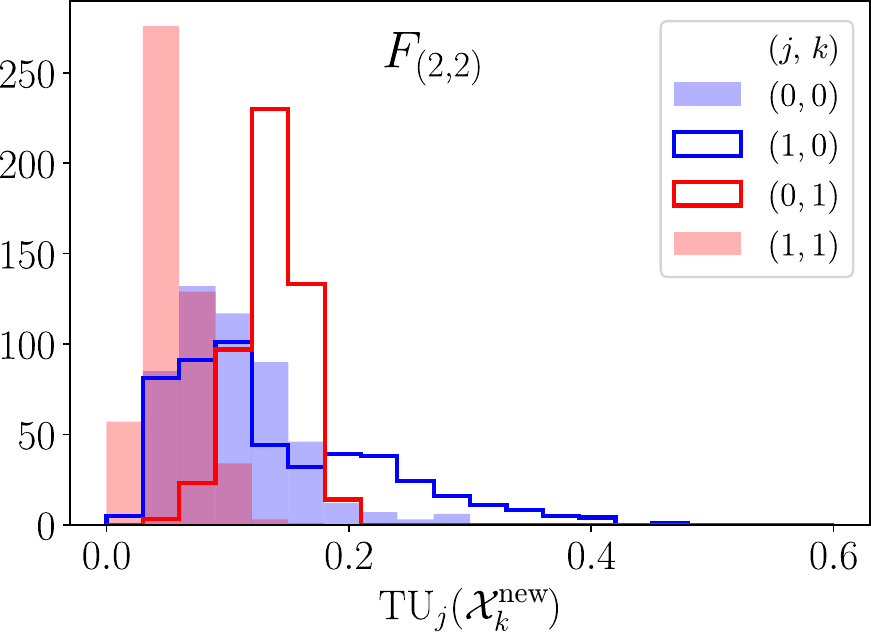}
  \hspace{0.5em}
  \includegraphics[width=0.48\textwidth]{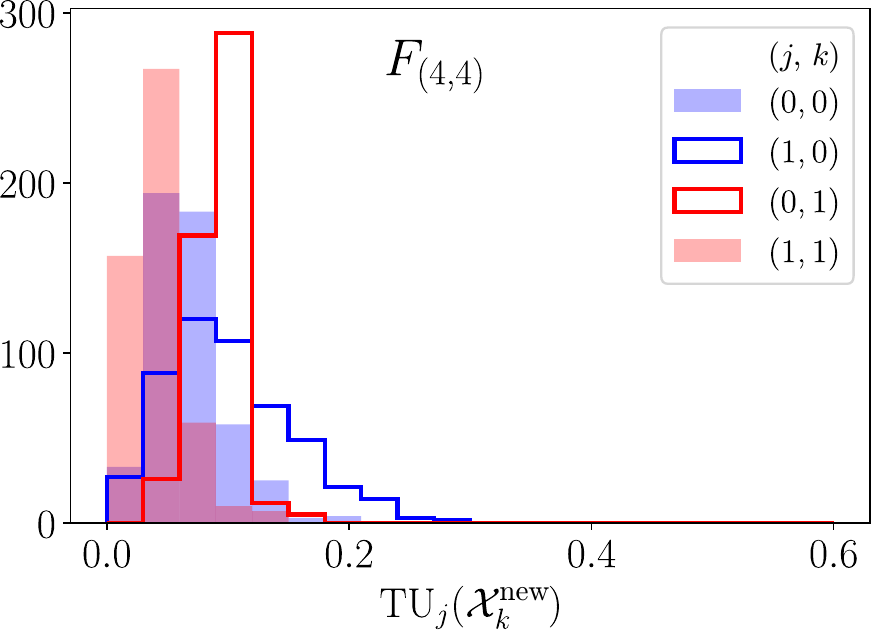}
  \vspace{0.5em}\\
  \includegraphics[width=0.48\textwidth]{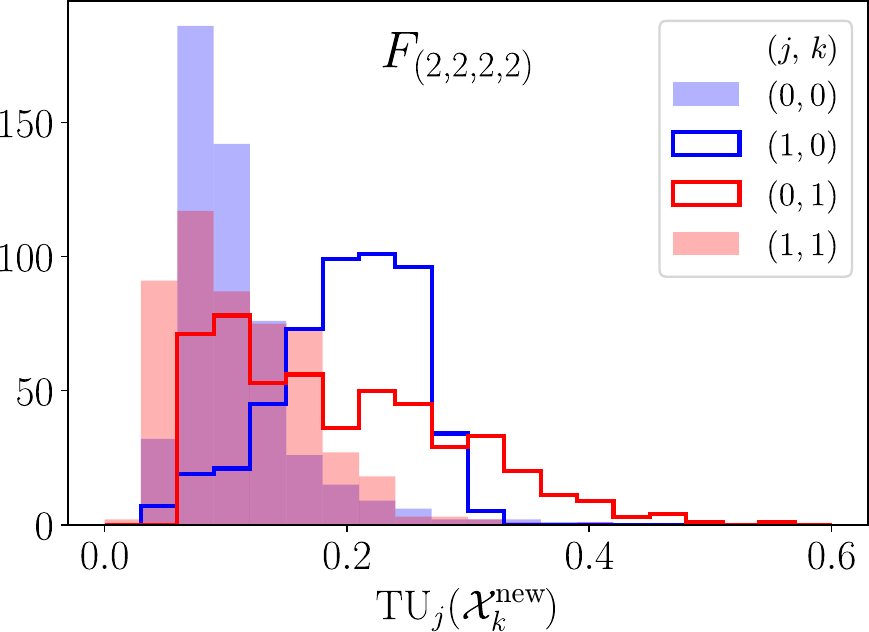}
  \hspace{0.5em}
  \includegraphics[width=0.48\textwidth]{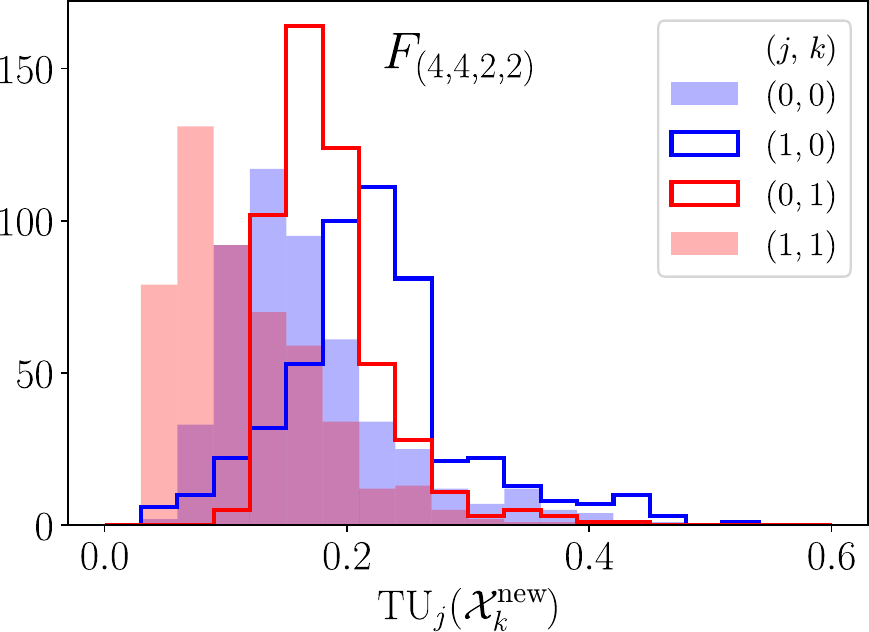}
  \caption{Histogram of $\TU_j(\calX^\text{new}_k)$ (i.e., $\TU_j(\bX_k)$ with $\bX_k\in \calX^\text{new}_k$) evaluated by $F_{(2,2)}$ (left-top), $F_{(4,4)}$ (right-top), $F_{(2,2,2,2)}$ (left-bottom), and $F_{(4,4,2,2)}$ (right-bottom) with $\delta = 0.5$.  The data are labeled by $(j,k)$ as indicated in the legend.  The blue (and red) color is used for the $k=0$ (and $k=1$, respectively) data.  The filled (and blank) bars are used for the $j=0$ (and $j=1$, respectively) data.}
  \label{fig:TU_hist_del05}
\end{figure}

It is not straightforward to understand the reason why the performance of $F_{(2,2,)}$ is exceptionally damaged.  Nevertheless, for the intuitive insight, it is helpful to visualize more differential information as shown in Fig.~\ref{fig:TU_hist_del05}.  These panels in Fig.~\ref{fig:TU_hist_del05} show the histograms of $\TU_j(\calX^\text{new}_k)$ calculated from $\calD^\text{new}_k$ with the label $k$ for the tolerance $\delta = 0.5$.

The histogram is quite intriguing for us to grasp the overall features of the TU performance.  We would recall that the TU value should be small if the uncertainty is reduced.  Thus, for the successful judgment of correct and incorrect inference, it is desirable to have small values of $\TU_j(\calX^\text{new}_k)$ for $(j,k)=(0,0)$ and $(1,1)$ as compared to $(0,1)$ and $(1,0)$.  Besides, the TU essentially measures the distance between the distribution based on $\calD^\text{new}_k$ and the Fr\'{e}chet mean $\bar{\mu}_{k \in \{0,1\}}$ as in Eq.~\eqref{eq:TUk}.
The performance is expected to be good when the shape of the color-filled histograms is narrowly peaked and closer to zero significantly than the shape of the blank-opened ones.
For example, let us consider $R_{k=0}$ for $F_{(2,2,2,2)}$ and $F_{(4,4,2,2)}$.
The filled [$(j,k)=(0,0)$] and open [$(j,k)=(1,0)$] blue histograms in two bottom panels in Fig.~\ref{fig:TU_hist_del05} are well peaked and separated, so that the same level of performance is foreseen from them.  Indeed, $R_{k=0}$ takes similar values for $F_{(2,2,2,2)}$ and $F_{(4,4,2,2)}$.
The situation is different for $R_{k=1}$;  the filled [$(j,k)=(1,1)$] and open [$(j,k)=(0,1)$] red histograms are not localized nor separated for $F_{(2,2,2,2)}$.  This observation is attributed to the bad performance of $F_{(2,2,2,2)}$ for $R_{k=1}$.  Although the histogram plots are useful to diagnose performance, we need to understand what causes such qualitative differences between $F_{(2,2,2,2)}$ and $F_{(4,4,2,2)}$.

\section{Conclusions}

We investigated the use of Topological Uncertainty (TU), i.e., a topological data analysis–based measure constructed from trained feedforward neural networks (FNNs), as a diagnostic for Anomaly Detection in the inference of Neutron-Star equations of state (EoSs).  By introducing labels defined through a tolerance parameter, we distinguished successful and unsuccessful inferences for the purpose of quantifying the performance, and employed the TU to characterize trained-network responses in the  hidden layers.

Our numerical experiments demonstrated that the TU can effectively separate normal from anomalous outputs, with detection rates exceeding $90\%$ for suitably chosen network architectures and tolerance values.  Our proposed method is robust against moderate variations of hyperparameters and provides complementary information to conventional output-based measures for uncertainty quantification.  Specifically, the TU captures structural differences in hidden-layer activations even when the outputs appear deceptively confident.  Moreover, an advantage of our method is that, to evaluate this uncertainty measure, what is necessary is just the trained FNN, and there is no need to augment network architectures.

The present study constitutes, to our knowledge, the first successful application of the TU to Anomaly Detection in a concrete physics problem.  Although further work is needed to establish quantitative reliability guarantees and to explore broader classes of network architectures, the TU emerges as a versatile and computationally inexpensive post-hoc tool to improve the robustness of data-driven inference of dense matter EoSs and similar problems in general physics.

Looking ahead, the TU offers promising applications across physics;  the FNNs have successfully been applied to various problems, and the TU can be calculable once the FNNs are reinterpreted as mappings to the classifiction problem.  In gravitational-wave analyses, the TU could enhance parameter estimation by flagging unreliable waveform mappings.  For Neutron-Star x-ray observations, it may help assess the robustness of mass–radius inference pipelines against systematic biases (see Ref.~\cite{Brandes:2024vhw} for a direct attempt).  In heavy-ion collision phenomenology, the TU could serve as a safeguard for surrogate networks extrapolating beyond the trained regime.  Finally, in numerical lattice-QCD studies, the TU may provide a new diagnostic to identify potentially insecure extrapolations from limited or noisy data.  These applications would highlight the TU as a physics-informed tool for Anomaly Detection and reliability assessment, with potential to uncover new phenomena in data-driven physics.

\begin{acknowledgments}
We thank Yuichi~Ike for useful discussions and insightful lectures about Topological Data Analysis and Topological Uncertainty, which inspired the authors to initiate this work.
We also thank Yuji~Hirono, Tatsuhiro~Misumi, and Ken~Shiozaki for having discussions on related topics regularly.
The authors are supported by JSPS KAKENHI Grant Nos.~22H05118 (K.F.\ and S.K.) and 25K07298 (S.K.).
\end{acknowledgments}


\bibliographystyle{apsrev4-1}
\bibliography{persistence_eos_master_plus}

\end{document}